\documentclass[letterpaper, 10 pt, conference, final]{ieeeconf}  

\IEEEoverridecommandlockouts                              

\usepackage{array,multirow}
\usepackage{hyperref}
\usepackage{tabularx}

\usepackage{bmpsize}
\usepackage{amsmath}
\usepackage{pifont}
\usepackage{amssymb}

\usepackage[final]{changes}
\definechangesauthor[name={Karla}, color=teal]{KS}
\definechangesauthor[name={Rado}, color=orange]{RS}
\definechangesauthor[name={Petr}, color=brown]{PV}

\title{\LARGE \bf
Tell and show: Combining multiple modalities to communicate manipulation tasks to a robot}

\author{
Petr Vanc$^{1}$
\and
Radoslav Skoviera$^{1}$
\and 
Karla Stepanova$^{1}$
\thanks{$^{1}$Czech Technical University in Prague, Czech Institute of Informatics, Robotics, and Cybernetics, \texttt{petr.vanc@cvut.cz}, \texttt{radoslav.skoviera@cvut.cz}, \texttt{karla.stepanova@cvut.cz}}
}

\begin{document}

\maketitle
\thispagestyle{empty}
\pagestyle{empty}

\begin{abstract}
As human-robot collaboration is becoming more widespread, there is a need for a more natural way of communicating with the robot. This includes combining data from several modalities together with the context of the situation and background knowledge. Current approaches to communication typically rely only on a single modality or are often very rigid and not robust to missing, misaligned, or noisy data. 
In this paper, we propose a novel method that takes inspiration from sensor fusion approaches to combine uncertain information from multiple modalities and enhance it with situational awareness (e.g., considering object properties or the scene setup). We first evaluate the proposed solution on simulated bimodal datasets (gestures and language) and show by several ablation experiments the importance of various components of the system and its robustness to noisy, missing, or misaligned observations. Then we implement and evaluate the model on the real setup. 
In human-robot interaction, we have to also consider if the selected action is probable enough to be executed or if we should better query humans for clarification. For these purposes, we enhance our model with adaptive entropy-based thresholding that detects the appropriate thresholds for different types of interaction showing similar performance as fine-tuned fixed thresholds.  



 \end{abstract}

\maketitle

\section{Introduction}

Human communication relies on combined information from several modalities such as vision, language, gestures, eye gaze, or facial expressions. 
These individual modalities support and complement each other, allowing humans to navigate missing, noisy, or unclear information and detect misaligned (conflicting) signals. Additionally, humans take into account the context of the situation and background knowledge, such as appropriate objects for the given action, enhancing the efficiency and robustness of communication.


On the contrary, current human-robot interaction setups usually facilitate rigid communication. They either rely solely on one modality (e.g., language \cite{pires2005robot}, or gestures \cite{Vanc2023}) or employ different modalities to specify individual message components in very strict scenarios. In the second scenario, language is primarily used to specify the type of action. Other modalities (e.g., eye-gaze or gestures) may assist in determining values for individual parameters, such as the target location (e.g., "Glue the bolt here")~\cite{behrens2019specifying} or other action-related parameters like direction (e.g., 'Move in this direction') or the type of movement~\cite{yongda2018research}). 

Despite efforts to merge information from various modalities, existing approaches often do so in a naive manner~\cite{1045664}. To foster more natural human-robot collaboration, a more general approach is needed to merge information from diverse data sources and accurately determine human intent. This approach should be capable of deciding when to be confident about the detected intent and when to seek clarification or repetition from the human.


Works in the area of sensor fusion \cite{bordes2013information,xu2016multimodal} present approaches for combining uncertain data from various sources. To be applicable in human-robot interaction scenarios, these approaches need to be extended to take into account also feasibility of individual actions and their parameters. This includes assessing whether the robot can reach or pick up an object and if the object can serve as a container.

\begin{figure}
  \centering
  \includegraphics[trim={0 0.2cm 0 0.5cm},width=1.0\linewidth]{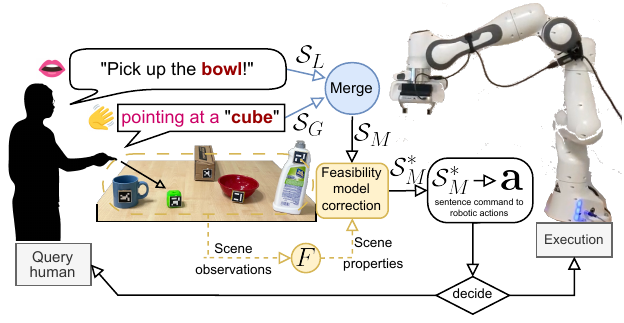}
  \caption{Human-Robot Interaction experimental setup. The user's speech is captured by the microphone and the hand is captured by a hand detection device (e.g. Leap Motion Controller \cite{Weichert_Bachmann_Rudak_Fisseler_2013}).}
  \label{fig:mm_intro}
  \vspace{-1.5em}
\end{figure}



In this paper, we take inspiration from the areas of information theory and sensor fusion \cite{smarandache2006depth} and propose a merging algorithm that provides a robust, context-aware solution for combining information from various modalities (see Fig.~\ref{fig:mm_intro}). Our approach handles multiple beliefs over possible actions from different modalities, updating the probability of these actions and their parameters by simultaneously combining information and checking the feasibility of the given combination in the current scenario. Furthermore, we utilize cross-entropy measures to evaluate the information conveyed in different modalities and use it to weigh the data sources. 

For effective human-robot interaction, it is crucial to decide whether to execute the most probable action or to seek clarification or repetition from the user. Executing a wrongly detected action could lead to significant issues. Therefore, we propose and statistically evaluate an entropy-based automated thresholding mechanism to determine the most appropriate interaction mode in human-robot scenarios.

We evaluate our proposed approach using well-controlled artificial data from gesture and language modalities, incorporating varying levels of alignment, noise, scene complexity, and action types. Some datasets contain data with conflicting information in both modalities, only resolvable based on context. Multiple ablation studies highlight the importance of different parts of the system.





In summary, the main contributions of the paper are:
\begin{itemize}
    \item A context-aware model for merging data from multiple modalities that is robust to noisy and misaligned data.
    \item Enrichment of the model through entropy-based automated thresholding to decide on the interaction mode in human-robot scenarios. 
    \item A thorough evaluation of the proposed model on simulated and real world dual-modality (gestures and language) datasets with varying complexity, including several ablation studies.
\end{itemize}

Datasets, code, and models are on the project site
\footnote{\label{projectwebsiteref}Project website:
\href{http://imitrob.ciirc.cvut.cz/mergingModalities.html}{http://imitrob.ciirc.cvut.cz/mergingModalities.html}}. 

\section{Related work}


Numerous studies have delved into the integration of multiple modalities, particularly gestures and language, within the scope of human-robot interaction. However, these efforts often relegate gestures to a secondary role, where language primarily dictates the action to be taken and gestures are used to provide additional details such as trajectory, distance, movement type~\cite{yongda2018research}, or position~\cite{behrens2019specifying, Du2018Online}. Works of Holzapfel~\cite{holzapfel2004implementation} and A.~Ekrekli~\cite{ekrekli2023co} focus on the fusion of information from pointing gestures and verbal commands to identify the intended object. Yet, to truly leverage the knowledge of data available from diverse modalities, it is desired to value all modalities equally across all aspects of communication, including action specification. A notable limitation of current methodologies is their restricted scalability, typically confining to no more than two modalities and overlooking contextual information.
The task of integrating diverse information sources, including vision, language, and gestures, to refine robot control and decision-making, has been thoroughly surveyed by P.~Atrey~\cite{Atrey2010MultimodalFF} in the context of multimedia analysis. 
In~\cite{smarandache2006depth}, the authors explore diverse methodologies for integrating uncertain data, while another study~\cite{JOSANG2010192} evaluates various belief conjunction strategies, highlighting their application in edge cases. Our research intersects significantly with sensor fusion~\cite{44007, bordes2013information,xu2016multimodal}, focusing on fusing inputs from a range of sensors across different communication modalities. Yet, our approach diverges by integrating fusion directly with decision-making processes, dealing with more complex attributes of the ``classes'' involved, such as the arity and properties of actions, or properties of the available objects, enriching the fusion process. This diverse application of sensor fusion shares conceptual similarities with its use in medical image segmentation scenarios~\cite{huang2022application}, underlining the versatility and challenges of integrating multi-source data.

The final part of our system hinges on its adeptness at contextualizing within the environment, a capability underpinned by the application of Recursive Bayesian Inference~\cite{8593766}. This methodology is the key to assimilating and interpreting varied data streams to deduce the most probable outcomes or states. Notably, contemporary research in this domain frequently focuses on leveraging controllers to fuse observations—predominantly from joystick maneuvers. This approach, however, contrasts with our system's broader scope, which incorporates a richer array of inputs, including gestures and verbal commands.

\section{Problem formulation}
\label{sec:problem_formulation}

The problem we are solving is to determine the intended manipulation action and its parameters based on combined information from different modalities while taking into account the context of the situation. First, we specify our assumptions and then provide the formulation of the problem.

\noindent \textbf{Human intent.} We define human intent $I$ as a triple $I = (ta, {to}, \mathbf{ap})$, where $ta$ is the target action ($ta \subset \mathcal{A}$, where $\mathcal{A}$ is a set of all available actions), $\mathbf{to}$ is a target object to be manipulated with  ($to \subset \mathcal{O}$, where $\mathcal{O}$ is a set of available objects) and $\mathbf{ap}$ is a list of other auxiliary parameters (e.g., storage location/object, distance, etc.). For example, human intent "pour the cup to bowl from 5 cm height" can be expressed as: \textit{I = (pour, [cup], [storage = bowl, h=5cm]}).

\noindent \textbf{Action parameters.} We expect that each action $a \in \mathcal{A}$ has $K$ compulsory and $L$ voluntary parameters ($K, L \in \{0,...,n\}$) each of them having a given and unique category 
(we call this \textit{signature} of the action). We distinguish, for example, parameters of the category action {$C_a$}, manipulated object {$C_o$}, storage object {$C_s$}, distance {$C_d$},  direction {$C_a$}, quantifier (amount), etc.. One action can contain a maximum of one parameter of the given category. For example, the above-mentioned action \textit{pour} has 2 compulsory parameters (target object of category {$C_o$} and container object of category {$C_s$}) and 2 voluntary parameters (height from which to pour of category {$C_d$} and angle under which to pour of category \textit{angle}). Considering this, the human intent has to be specified by at least $K+1$ information units (target action and all its compulsory parameters). 
Note that it is sufficient to specify the compulsory parameters only in one of the modalities. 

\noindent \textbf{Object features.} Let's suppose that the current scene contains a set of objects $\mathcal{O}_s \subset \mathcal{O}$. Each object $o \in \mathcal{O}$ has attributed a list of features $F_o$ (e.g., pickable, container, full, etc.): $F_o = \{f_1,...,f_F\}$, where $f_i \in \mathcal{F}$, $\mathcal{F}$ is a set of all available features. Correspondingly, each action $a$ has requirements on the features (properties) of the target object(s) (e.g., \textit{pour} action might require that $to$ is pickable, reachable, full and not glued and $so$ (storage object) has to be reachable and liquid-container). This can be noted as: $(\mathcal{F}_{to}\mid pour) = f_1 \& f_2 \& f_3 \& \neg f_6$, $(\mathcal{F}_{so}\mid pour) = f_1 \& f_4$.

\noindent \textbf{Observations from modalities.} Let's suppose we perceived information from $M$ modalities in the form of sentences $\mathbf{S}_1,..., \mathbf{S}_M$. Each sentence is composed of $m$ words $\mathbf{w}_i$. For each of these words we can estimate the probability that it is carrying information about the parameter of the given category (i.e., we can estimate values $p(\mathcal{C}(\mathbf{w}_i) = C^q), \forall C^q \in \mathcal{C}$). 
Each word $\mathbf{w_i}$ contains a likelihood vector where each value provides a likelihood of the given option. For example, if the user's intent to perform action \textit{wave} is expressed via gesture, then the sentence from gesture modality $\mathbf{S}_G$ contains one word $\mathbf{w}_1$ expressed by a likelihood vector with $P$ values: $\mathbf{w}_1 = (w_1^1,...,w_1^P$), where $w_1^i$ represents likelihood of the gesture $i$ (gesture mapped to action $i$), i.e., $w_1^i = l (g_i)$, and $P$ is the number of available gestures mapped to individual actions, i.e., $P= len\{\mathcal{G}\}$. In this case, the highest probability value will be observed for the gesture mapped to the action \textit{wave} (see video and project website for more examples).

Now we can formulate the task as follows:

\noindent \textbf{Problem formulation.} Determine the human intent $I$ (i.e., the target action $ta$ and its parameters) based on information received from various modalities in the form of sentences $\mathbf{S}_1,..., \mathbf{S}_M$ of variable length while taking into account the set of available objects $\mathcal{O}_s$ and their features $\mathbf{F}_o$ as well as requirements of individual actions $a$ such as number of parameters or required features of manipulated objects.

\section{Proposed solution}

\begin{figure}
  \vspace{0.4em}
  \centering
  \includegraphics[trim={0 0.0cm 0.0cm 0.0cm},width=0.8\linewidth]{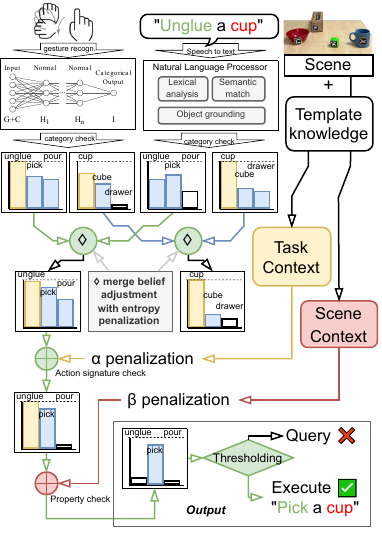}
  \vspace{-1.2em}
  \caption{Diagram of the proposed model for the case of two modalities (hand gestures and natural language) specifying action with one parameter (target object). Heard sentence ``Unglue a cup'' is correctly resolved into ``Pick a cup'' based on a fusion of data from both modalities and task and scene context".}
  \label{fig:model_diagram}
  \vspace{-1.8em}
\end{figure}

In this section, we will describe our proposed solution for the previously formulated problem, i.e., determining intent based on information from various modalities (see Fig.~\ref{fig:mm_intro} and for walkthrough example, see Fig.~\ref{fig:model_diagram}).

\subsection{Theoretical background / definitions}
\label{sec:background}
We define diagonal cross-entropy $DCH$ as cross-entropy between a vector of likelihoods $\boldsymbol{l}$ and a set of one-hot vectors $\boldsymbol{dv}_j, j \in \{1,..,N\}$, where $N$ is the length of $\boldsymbol{l}$. Each vector $\boldsymbol{dv}_j$ is constructed as follows:
\begin{equation}
dv_j(i) = 
    \begin{cases}
        0 & i \neq j \\
        1 & i = j
    \end{cases},
    i \in \{1,..,N\}.
\end{equation} 
The vector $\boldsymbol{dv}_j$ symbolizes an instance of a probability distribution over the same elements as in $\boldsymbol{l}$, where only the $j$-th element has a probability $1$ and all other elements have a probability of $0$. The $DCH$ is computed as follows:
\begin{equation}
\footnotesize
    \begin{split}
    DCH(\mathbf{l}) &= \boldsymbol{h}, \mathrm{where } \\
    h(j) &= H(\mathbf{l}, \boldsymbol{dv}_j), j \in \{1,..,N\}~,
\end{split}
\end{equation}

where $H$ is the information entropy. $\boldsymbol{h}$ then signifies the dissimilarity between $\boldsymbol{l}$ and the extreme case where the corresponding element in $\boldsymbol{l}$ would be detected with absolute certainty.

\subsection{Assumptions}
To simplify the notation without loss of generality, we assume a one-to-one mapping between action words and actions). 
For example, in the case of gestures, we expect to know the mapping $\mathcal{A} = \mathcal{M}_{\mathcal{G}}(\mathcal{G}_a)$ between action gestures $\mathcal{G}_a$ and individual actions $\mathcal{A}$, i.e., gesture $g_i$ corresponds to the action $a_i$. 
How learn more general mappings from observations was shown in our previous work~\cite{Vanc2023}.

Second, we assume a good synchronization of the data from different modalities. We expect the sentence of each modality to have the same length and the words that are missing in the given modality contain empty vectors (for example detected from pauses). This assumption is natural, as in human communication the information is typically expressed in a synchronized manner over all modalities. 


\subsection{Merging algorithm}

\label{sec:merging}
First, we describe the main merging algorithm which is utilized to merge input data (sentences $\mathbf{S}^1,...,\mathbf{S}^M$) from individual modalities and create the fundamental merged multimodal sentence $\mathbf{S}^\mathcal{M}$. Assuming synchronized expression of information across modalities, all sentences share the same length ($n$) and $i$-th words across all modalities correspond. The words of the merged sentence are determined as follows:

\begin{equation}
    w^\mathcal{M}_i=  \Diamond_{m=1}^M\left(W_m*\mathbf{w}^m_i\right),\forall i \in [1,n],
    \label{eq:merge}
\end{equation}

where $\Diamond$ represents the mixing operation (we compare results for maximum, addition, and multiplication, but it can be also any other operator). The mixing goes over all available modalities $m$ ($M$ is the total number of modalities), where $W_m$ is the weight of the modality $m$, and $\mathbf{w}^m_i$ is the $i$-th word from the sentence $\mathbf{S}^m$. Please note that if information about the word $\mathbf{w}_i$ is not expressed in a given modality, then $\mathbf{w}^m_i$ is empty and is not considered in the merging. For example, if we have on the input language sentence "Give me cup" and gesture signalizing action "give me", then both sentences have length 2, but the gesture sentence has one empty word: $S^L$ = [$\mathbf{w}^L_1$,$\mathbf{w}^L_2$], $\mathbf{S}^G = [\mathbf{w}^G_1,[]]$ and merged sentence is then: $S^\mathcal{M}= [[W_L*\mathbf{w}^L_1\Diamond W_G*\mathbf{w}^G_1], W_L*\mathbf{w}^L_2]$. 

\subsubsection{Belief adjustment by entropy penalization}

We introduce entropy penalization, that determines how unlikely the measurement of action $a_j$ within modality $m$ is a true measurement rather than noise. To estimate this, we use the $DCH$ (see Sec.~\ref{sec:background}). 
In the case of using this penalization, the Eq.~\ref{eq:merge} would get one additional weighting term:
\begin{equation}
\footnotesize
        w^\mathcal{M}_i=  \Diamond_{m=1}^M\left(W_m*[\frac{1}{DCH(\mathbf{w}_m)}\odot\mathbf{w}_m]\right), \forall i \in [1,n],
\end{equation}
where $\odot$ signs element-wise multiplication.


\subsection{Adjusting sentence for feasibility - selection of action}
\label{sec:selAct}
After generating the merged sentence, the second step involves determining, which specific action, along with its corresponding parameters, is most likely expressed by the human. This step takes into account the observations of the scene and the human. Initially, we compute likelihoods for all possible actions and identify the most probable action. 


 Given that we for each word know the category of the parameter, we know that target action is expressed by word $\mathbf{w}^\mathcal{M}_{C_a}$ from the merged sentence (computed based on Eq.~\ref{eq:merge}) and we can compute the likelihood of each action $a_j \in \mathcal{A}$ as follows:
\begin{equation}
    l(a_j)=  \mathbf{w}^\mathcal{M}_{C_a}(j) * \alpha_{a_j} * \beta_{a_j}, 
    \label{eq:actionlik}
\end{equation}
where the $\alpha$ is the penalization parameter for not fitting action parameters (see section~\ref{sec:penal_arity}) and $\beta$ is the penalization parameter for object features that are not satisfied (see section~\ref{sec:penalization_props}).

The most probable action $a_{j^*}$ is then selected by:
\begin{equation}
    j^* = argmax_{j}(l(a_j))
    \label{eq:argmax}
\end{equation}
We execute this action only if it exceeds the given threshold and has enough difference from other options (see Sec.~\ref{sec:thresholding})



\subsubsection{Penalization for unaligned parameters}
\label{sec:penal_arity}

Let's $\mathcal{C}$ be the set of all available parameter types ($\mathcal{C} = \{C_1,...,C_n\}$). Given action $a$ has compulsory parameters $\mathcal{C}^c \subset \mathcal{C}$ and voluntary parameters $\mathcal{C}^v \subset \mathcal{C}$, $A$ being a penalization parameter, and that merged sentence $\mathbf{S}^\mathcal{M}$ over all modalities is containing specification for parameters $\mathcal{C}^w \subset \mathcal{C}$, we can compute penalization parameter $\alpha$ as follows: 
\begin{equation}
\footnotesize
    \begin{split}
    L &= \sum_{C^c_i \in \mathcal{C}^c} (1-p(C^c_i \in \mathcal{C}^w)) + \sum_{C_i \in \mathcal{C}\mid C_i \notin \mathcal{C}^c,\mathcal{C}^v} p(C_i \in \mathcal{C}^w) \\
    \alpha_a &= A^L
\end{split}
\end{equation}
This means that we increase the penalization for every parameter that is compulsory and not available in the observed sentences (first term) and we also increase the penalization for every parameter that is in the observed sentences but is neither a compulsory nor voluntary parameter 
(second term). 

\subsubsection{Penalization for unaligned object properties}
\label{sec:penalization_props}
The second penalization is considering the scene context, i.e., it penalizes non-aligned object properties in the incoming sentences with the requirements of the action. This penalization is only applicable for actions that have as a parameter either a target or storage object. Let's suppose that action $a_j$ has the following requirements on the target object properties: $\{\mathcal{F}(to)\mid a_j\} = f_1 \& f_2 \& f_3 \& \neg f_6$ (e.g., action \textit{pour} requires the manipulated object to be \textit{reachable, pickable, full} and \textit{not glued}). Let's expect that $\mathbf{f}^{o_i}$ is a vector of likelihoods of given features for the object $o_i$ (i.e., $f^{o_i}_6 = 0.2$, means that the likelihood that object $o_i$ is glued is 0.2). The misalignment $f^a_k$ between requirements of action $a_j$ on the target object ($f_k$) and actual property of target object $o_i$ ($f_k^{o_i}$) can be then computed as:
\begin{equation}
\footnotesize
    \forall k, \mathrm{where }f_k \in \{\mathcal{F}(to)\mid a_j\}: f^a_k = abs(f_k^{o_i} - f_k),
\end{equation}
For example if $\mathbf{f}^{o_i} = (1, 0.8, 1, 0, 1, 0.2)$, then $\mathbf{f}^a = (0,0.2, 0, -, -, 0.2)$. 
We compute the final alignment of object $o_i$ with requirements to action $a_j$ as $A(o_i, \mathcal{F}(to\mid a_j) = 1-max(\mathbf{f}^a)$ (i.e., in our case $A(o_i, \{\mathcal{F}(to)\mid a_j\}) = 0.8$). 

Finally, the penalizing parameter $\beta$ is computed as follows:
\begin{equation}
\footnotesize
\begin{split}
    \beta = &\max \{A(o_i, \mathcal{F}(to\mid a_j))\} * \max \{A (o_k, \mathcal{F}(so\mid a_j))\},\\
    &\forall o_i: w^{\mathcal{M}}_{C^{to}}(o_i)>T_{clear}, \forall o_k: w^{\mathcal{M}}_{C^{so}}(o_k)>T_{clear}.\\
\end{split}
    \end{equation}
This means that we find maximal alignment among all target objects whose probability after merging ($w^{\mathcal{M}}_{C^{to}}(o_i)$) exceeds the threshold for a clear option $T_{clear}$ (thresholds are either fixed or determined automatically based on entropy, see Sec.~\ref{sec:thresholding}). The same is computed also for storage objects and these two values are multiplied. This means that there is a fully fitting target object with enough high probability, the parameter $\beta$ is 1 and the action $a_j$ is not penalized. On the contrary, if there is no fitting object among the object with probability exceeding the threshold, $\beta = 0$, and the action is discarded. Please note, that if the action has among compulsory parameters only one of the target or storage objects, the other term is automatically omitted from the computation. If the action requires neither of them, the penalization is not considered at all.



\subsection{Selecting other action parameters}

When we select the target action $a_{j^*}$ (see Eq.~\ref{eq:argmax}), we still have to determine its specific parameters. For each of the parameters required by the action ($\mathcal{C}^c$) we compute $argmax$ over all the available options exceeding the noise threshold and parameter value $k^*$ is selected: $\forall C_i \in \mathcal{C}^c: k^* = argmax_k(w_{C_i}^\mathcal{M}(k))$. For example, if the action $a_{j^*}$ requires as compulsory parameter target object and if the word expressing parameter target object ($w^\mathcal{M}_{C_{o}}$) contains likelihoods of objects $o_1$ and $o_4$ that exceed the threshold $T_{clear}$ (see Sec.~\ref{sec:thresholding}), and $l(o_4)>l(o_1)$, object $o_{k*}=o_4$ will be selected as the resulting value for target object. 

If for some of the parameters, none of the options exceeds the given threshold, action is not executed and the user should be prompted for clarification (same as in the case of unclear information about action). 

\subsection{Decision over the actions for interaction mode}
\label{sec:thresholding}

We distinguish three types of interaction modes based on the clarity of the incoming information: i) \textit{clear} information -- action will be executed, ii) \textit{unclear} information -- action will not be executed, a user is prompted for more information, and iii) \textit{noise} -- command not sufficiently recognized (user will be asked for repeating the input).

To decide the merged probability of each action which case it belongs to, we use the following two approaches.

\subsubsection{Entropy-based threshold}

In this approach, we first compute $\boldsymbol{h} = DCH\left(\mathbf{l}\right))$. Then, for each action, we decide:

\begin{equation}
    d\left(a_{i}\right)
    \begin{cases}
        clear & \boldsymbol{h}\left(i\right)>t_{E}\\
        unclear & t_{n}<\boldsymbol{h}\left(i\right)\leqq t_{E}\\
        noise & \boldsymbol{h}\left(i\right)\leqq t_{n}
    \end{cases}
\end{equation}

The noise threshold $t_n$ is a very small number, which we based on the evaluation of the real signals set to $0.05$. The threshold $t_E$ is computed automatically. We tested two values for threshold $t_E$. The value $t_E=H\left(\mathbf{l}\right)$ gives reasonably good results and should work well in general. However, we found that better results can be achieved if $t_E$ is set to the average entropy of a vector of length equal to $\mathrm{len}(\mathbf{l})$, sampled from a uniform distribution. This can be seen as comparing against "a white noise" output from the detector. If there are multiple clear actions, or if all the actions are unclear, we query the user for verification.

\subsubsection{Fixed probability threshold}
\label{sec:thresholds_fixed}
In this case, we introduce fixed thresholds $t_{C}$ and $t_{U}$ to classify between noise, unclear, and clear action. 

\section{Experimental setup}

\begin{figure}
    \vspace{0.4em}
    \centering
        \includegraphics[width = 0.23\textwidth]{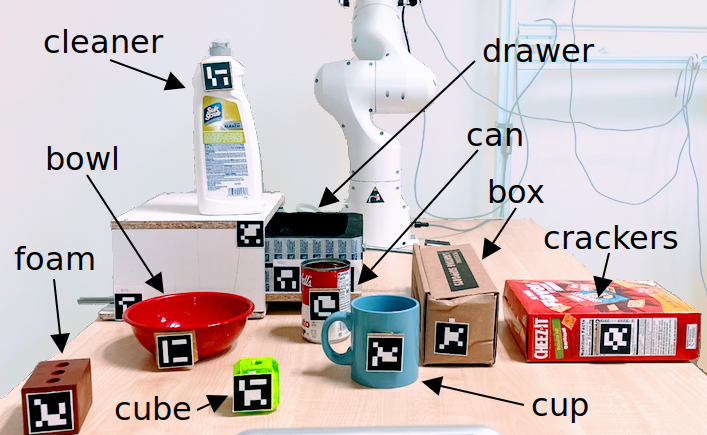}
    \includegraphics[width = 0.23\textwidth]{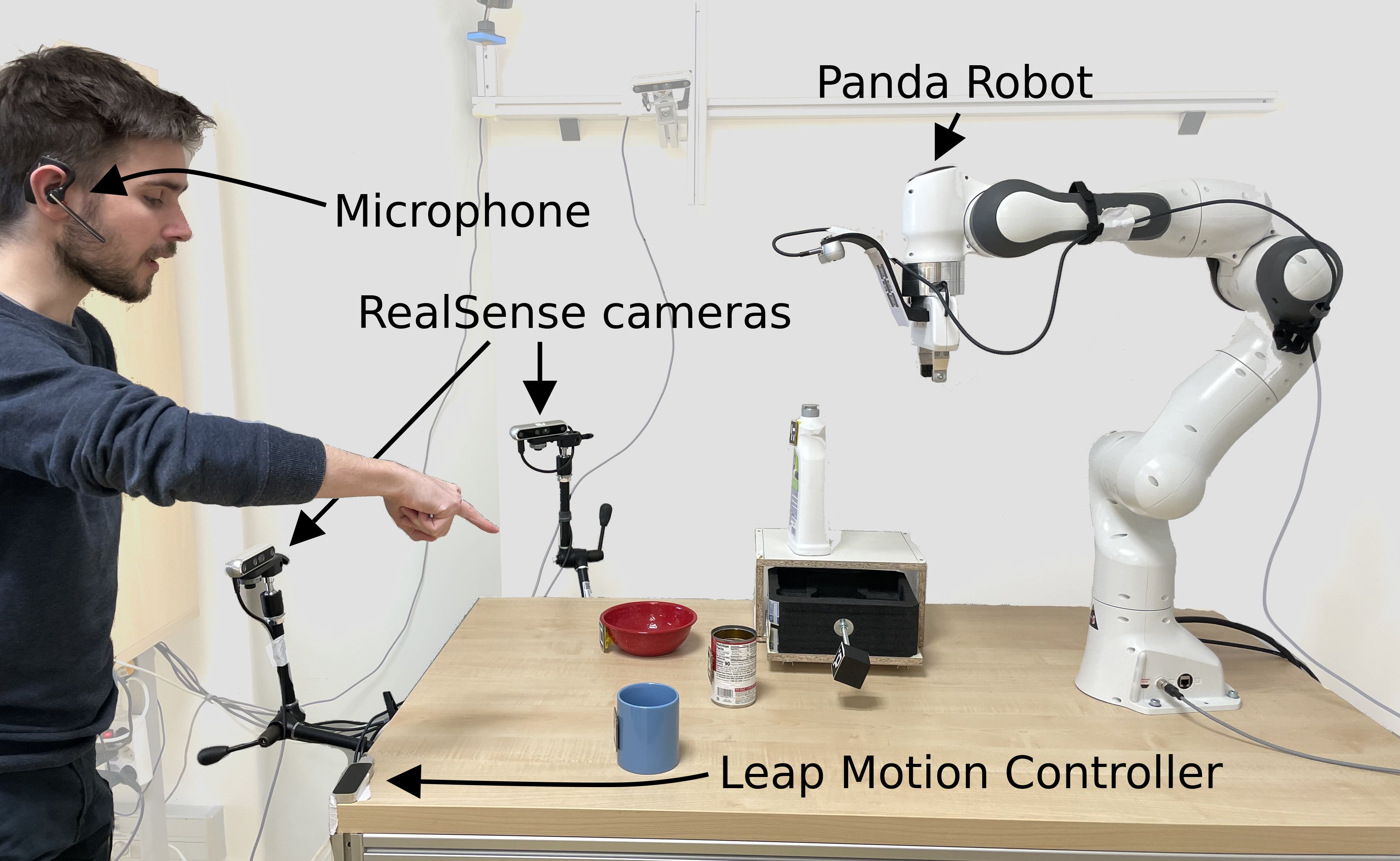}
    \caption{Real experimental setup. (left) Set of all objects used in the real experiment. (right) Example of the setup with 3 objects (box, can, and cleaner) and two storage areas (drawer, bowl) for instructions \textit{"Put the can into the drawer"}. See the attached video and project website\textsuperscript{\ref{projectwebsiteref}} for more examples.}
    \label{fig:real_setup}
    \vspace{-1em}
\end{figure}

We consider tabletop scenarios overlooked by a Realsense D455 RGBD camera with a Franka Emika Panda robot manipulator and objects spawned randomly on the table (see Fig.~\ref{fig:real_setup} (right)). For the real experiment, the positions of the objects are detected using the attached AruCo markers and updated in real-time. In both simulated and real experiments, we considered two input modalities (gestures and language) for instructions. For the real experiment, gestures are detected using a Leap motion sensor attached to the corner of the table and recognized using our Gesture toolbox~\cite{gesturetoolbox} (see more details in Sec.~\ref{sec:real_inputs}).   

\subsection{Assumptions for the experiment}
\label{sec:exp_assumptions}
To be able to focus fully on the evaluation of the context-aware merging of modalities, we make in our experiment the following assumptions: 1) we assume that object properties are just binary values (i.e., object is or isn't reachable), 2) we expect to know the category of each word (Sec.~\ref{sec:selAct}), and 3) actions have only compulsory and no voluntary parameters.

We used the following settings for our experiments. Parameter $A$ (Sec.~\ref{sec:penal_arity}) is set to $0.2$, for all experiments. 
Fixed thresholds (Sec.~\ref{sec:thresholds_fixed}) were optimized for the simulated experiment and set to: $t_C = 0.25$ and $t_U = 0.11$.

\subsection{Set of actions and objects}
\label{sec:actions}

We predefined a set of actions with 0, 1 or 2 compulsory parameters: 3 actions with no parameter (\textit{move up}, \textit{stop}, \textit{release}), 3 actions with 1 parameter (\textit{pick up X}, \textit{unglue X}, \textit{push X}), and 3 actions with 2 parameters (\textit{pour} A to B, \textit{put A into B}, \textit{stack A on B}). For our experiments, we used the following set of object and storage properties: $\mathcal{F}=$\{\textit{glued}, \textit{pickable}, \textit{reachable}, \textit{stackable}, \textit{pushable}, \textit{liquid container}, \textit{(basic) container}, \textit{full-stack}, \textit{full-liquid}, \textit{full-container}\}. Each action has specified requirements for the target object and storage object (e.g., pour action requires that the target object is reachable, pickable, full, and not glued and that the storage object is reachable and liquid-container).

For the simulated experiment, we generate a scene by distributing on the table a set of 5 random objects and 3 storage objects. Their properties are set according to the given dataset (see Sec.~\ref{sec:datasets}). 

For the real experiment, we used a set of 5 objects (can, cup, cube, box, and cleaner) and 4 storage objects (drawer, foam, crackers, and bowl) (see Fig.~\ref{fig:real_setup} (left) for the set of all objects used in the real experiment). For each task, $3$ objects and $2$ storage objects are distributed on the table (see Fig.~\ref{fig:real_setup} (right) for the real setup example). Each object has some assigned properties (e.g., liquid-container, pickable, etc.) and variable properties (e.g., reachable, glued, full-liquid, etc.) (see our website\textsuperscript{\ref{projectwebsiteref}} for details). The variable properties of the objects and storage objects, as well as the corresponding gesture and language commands, are set according to the given dataset (see Sec.~\ref{sec:datasets})).

\subsection{Gesture and language instructions processing}
\label{sec:real_inputs}

In the real experiment, for gesture detection, we utilized a Leap Motion sensor attached to the corner of the table, which captures the bone structure of the hand in real-time (see Fig.~\ref{fig:real_setup} (right)). The data from the Leap Motion sensor are processed in our Gesture toolbox~\cite{gesturetoolbox} where individual gestures are recognized. As described in Sec.~\ref{sec:actions} for the real experiment we consider human intent (see Sec.~\ref{sec:problem_formulation}) consisting of target action ($ta (C_a)$) and depending on the type of the target action optionally also of the target object ($to (C_o)$) and/or storage object ($so (C_s)$). The gesture sentence is recorded while the human holds a hand above the Leap Motion sensor. Individual gestures (i.e., words of the gesture sentence) are detected using cumulative evidence for the given gesture. Detected static and dynamic gestures are mapped to the 9 target actions used in the experiment. A pointing gesture activates the detection of target and storage objects. In this case, probabilities of individual objects are computed using the distance from the line from the pointing finger (using the approach described in~\cite{vanc2023communicating}). A closed hand separates individual words in the gesture sentence. Moving the hand away from the detection area finishes and sends the recorded gesture sentence. Each word, being a vector of probabilities, is assigned a category (e.g., a pointing gesture determines category $C_o$).  Refer to the attached video and project website\textsuperscript{\ref{projectwebsiteref}} for a real demonstration of the gesture setup. 

Language instructions in the real setup are transferred to text, 
parsed, filtered for filling words and synonyms, tokenized, and compared to individual language templates (see the available code at our project website\textsuperscript{\ref{projectwebsiteref}}). The final language sentence consists of words with one-hot activations. A constant noise is added to the zero values.

To ensure that the artificial data corresponds well to the real data, we generate probability vectors (simulating outcomes from gesture and language inputs) so they respect similarities observed in real data. To generate likelihoods of individual gestures in the gesture vector, we sample from the generated similarity table that has been created based on the sample dataset of real gesture data collected by our Gesture toolbox~\cite{gesturetoolbox}.
The similarity for the English language has been created based on Levenshtein distance of the phonological transcripts of the used words generated by the tool Metaphone \cite{McGreggor_2023}. See the available code at\textsuperscript{\ref{projectwebsiteref}}.

\subsection{Evaluated models, merge functions and thresholding}
\label{sec:models}

In our experiments, we perform ablation studies on the following models with varying types of penalization, merging function, and thresholding mechanisms.

\noindent \textbf{Penalization terms. }We compare the following basic models: 1) \textit{$M_1$ model}: Simple merge without any penalization term (i.e., Eq.~\ref{eq:merge}) without parameters $\alpha$ and $\beta$, 2) \textit{$M_2$ model}: introduces penalization $\alpha$ for unaligned parameters (i.e, Eq.~\ref{eq:merge} without parameter $\beta$), and 3) \textit{$M_3$ model}: a full model that introduces penalization for both unaligned parameters and object properties (i.e., the full model described in Sec.~\ref{sec:merging} and in Eq.~\ref{eq:merge}). See overview in Tab.~\ref{tab:models}. 

\noindent \textbf{Merge function. }We compare models using the following merge function in Eq.~\ref{eq:merge}: 1) \textit{maximum}, 2) \textit{multiplication}, and 3) \textit{addition}. We refer to the $M_1$ model with \textit{max} merging function as a \textbf{Baseline model}. 

\noindent \textbf{Thresholding. }Finally, we compare 1) \textit{entropy-based adaptive thresholding} vs. 2) \textit{fixed thresholds}. See an overview of the compared models and their variants in Tab.~\ref{tab:models}.


\begin{table}[ht]
  \caption{Ablation studies over models and merge functions.}
  \label{tab:models}
\begin{center}
\vspace{-0.4cm}
\scalebox{0.83}{
  \begin{tabular}{|c|c c c|c|c||c c|}
    \hline
                                  \multirow{2}{*}{Model}     & \multicolumn{3}{c|}{Merge function} & Signature\multirow{2}{*}{} & Property\multirow{2}{*}{} & \multicolumn{2}{c|}{Thresholding}\\
               &   $max$  & \textit{mul} ($\cdot$) & \textit{add} ($+$)        & pen. ($\alpha$) & pen. ($\beta$) & Fixed & Entropy \\ 
 \hline
             Baseline   & \color{teal}{\checkmark}   &  \color{red}{\ding{55}} & \color{red}{\ding{55}} & \color{red}{\ding{55}} & \color{red}{\ding{55}} & \color{red}{\ding{55}} & \color{red}{\ding{55}} \\
                  $M_1$  & \color{teal}{\checkmark} & \color{teal}{\checkmark}   & \color{teal}{\checkmark}    & \color{red}{\ding{55}} & \color{red}{\ding{55}} &\color{teal}{\checkmark} & \color{teal}{\checkmark}\\
 $M_2$   & \color{teal}{\checkmark} & \color{teal}{\checkmark} &\color{teal}{\checkmark}   & \color{teal}{\checkmark}   & \color{red}{\ding{55}} & \color{teal}{\checkmark} & \color{teal}{\checkmark}\\
 $M_3$   & \color{teal}{\checkmark} & \color{teal}{\checkmark} & \color{teal}{\checkmark}  &\color{teal}{\checkmark}   &\color{teal}{\checkmark} &\color{teal}{\checkmark} & \color{teal}{\checkmark}\\ \hline 
  \end{tabular}
  }
  \end{center}
  \vspace{-2em}
\end{table}


\subsection{Considered noise levels \texorpdfstring{$n$}{n}}
\label{sec:generating-noises}

\begin{figure}[ht]
  \centering
  \includegraphics[trim={0.0cm 0.00cm 0.0cm 0.0cm},clip,width=0.9\linewidth]{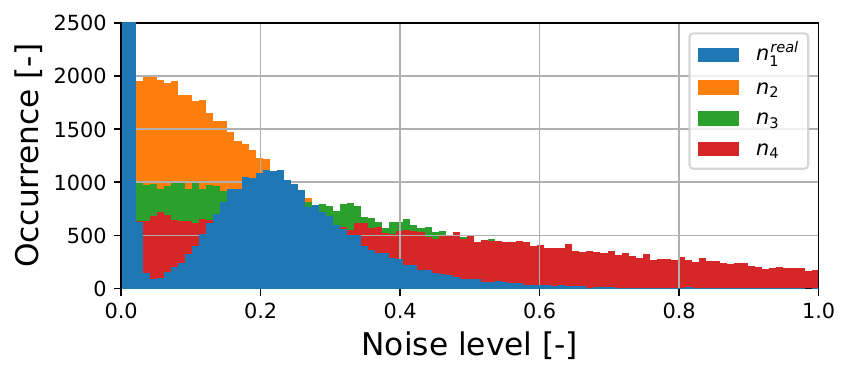}
  \vspace{-1.3em}
  \caption{Different levels of noise added to simulated data.}
  \label{fig:noisemodel}
  \vspace{-1em}
\end{figure}

For simulated experiments, we consider five noise levels (including no noise option) (see Fig.~\ref{fig:noisemodel}) that are added to the generated datasets (i.e., affecting likelihoods of detections): 
1) \textit{Zero noise} $n_0$ (used as a baseline), 
2) \textit{Real-data noise} $n_1$: Modeled based on the real data of gesture detections,
3) \textit{Regular noise} $n_2$: Artificial noise $\mathcal{N}(0.0, 0.2)$ with similar standard deviation as real-data noise,
4) \textit{Amplified noise $n_3$}: Artificial noise $\mathcal{N}(0.0, 0.4)$, and 5) - \textit{Extreme noise $n_4$}: Artificial noise $\mathcal{N}(0.0, 0.6)$.

\subsection{Simulated datasets \texorpdfstring{$D$}{D}}
\label{sec:datasets}

For our experiments, we generated the following simulated datasets, all including data from two modalities (gestures and language). We consider the following general cases that typically occur in human communication: 1) $\mathbf{D_{\mathcal{A}}^{sim}}$: Aligned dataset with full information available in both modalities (used as a baseline), 2) $\mathbf{D_{arity}^{sim}}$: Non-aligned dataset decisible based on the signature of the action (i.e., number and type of the parameters) (e.g., $S^L$ = ["pour a cup to bowl"], $S^G$ = ["pour a cup to right"] - feasible action is only "pour a cup to bowl" as direction parameter (right) is not a parameter of action pour), and 3) $\mathbf{D_{prop}^{sim}}$: Non-aligned dataset decisible based on the properties of the objects (e.g., $S^L$ = ["pour a cup to bowl"], $S^G$ = ["pour a cup to notebook"], feasible action is only pouring a cup to bowl, because the notebook is not a liquid-container). To generate the datasets, we first generate a random scene and select a random action from the 9 actions (see Sec.~\ref{sec:actions}) and its parameters. Afterward, we select alternative actions and/or parameters for the unaligned dataset and adjust the properties of the objects if necessary (see our website\textsuperscript{\ref{projectwebsiteref}} for an example of the generated dataset and the corresponding code). Finally, probability vectors simulating outcomes from gesture and language inputs are generated and additional noise is added optionally (see Sec.~\ref{sec:generating-noises}. 
Datasets $\mathbf{D_{arity}^{sim}}$ and $\mathbf{D_{prop}^{sim}}$ highlight the importance of the context of the situation, they are further combined into a single \textit{unaligned} dataset $\mathbf{D^{sim}_{\mathcal{U}}}$.

To ensure that the artificial data well corresponds to the real data, we use noise modeled based on the real-data (see Sec.~\ref{sec:generating-noises}) and we generate probability vectors (simulating outcomes from gesture and language inputs) so they respect similarities observed in real data (see Sec.~\ref{sec:datasets_with_specified_configurations}). To generate likelihoods of individual gestures in the gesture vector, we sample from the generated similarity table that has been created based on the sample dataset of real gesture data collected by our Gesture toolbox~\cite{gesturetoolbox}.
The similarity for the English language has been created based on Levenshtein distance of the phonological transcripts of the used words generated by the tool Metaphone \cite{McGreggor_2023}. 


\label{sec:datasets_with_specified_configurations}
In total, by combining $3$ types of dataset $D$ and $5$ noise levels $n$ we evaluated our models and different merging policies on $15$ datasets, each of 1000 samples. 


\subsection{Real datasets \texorpdfstring{$D^{real}$}{D real}}
\label{sec:datasets_real} 

We prepared 20 different setups for the real experiments, each featuring 2-3 objects and 2 storage objects (see an example in Fig.~\ref{fig:real_setup}). For each setup, we randomly selected the target action and corresponding target and storage object (if required for the given action) from the set of available objects while considering their fixed properties to ensure the target action's feasibility. Subsequently, we set variable properties of the objects (e.g., reachability) so that the target action remained feasible. The properties of other objects were set so that the target action was infeasible on them. Based on this, we prepared for each setup a corresponding valid instruction sentence expressing human intent (e.g., \textit{Put can to drawer}). For each of these setups, we also prepared an alternative instruction sentence with an unfeasible target action, which could be determined as false in half of the cases based on the signature of the action, i.e., type and a number of parameters (e.g., \textit{Pick the can in the drawer}, where \textit{pick} has only one compulsory parameter) and in the other half based on the properties of the objects (e.g., \textit{Pour the can in the drawer}, where the drawer is not a liquid container, so the can not be poured into it). We also prepared a second alternative sentence with a valid target action, but a different target object for which the target action is infeasible (e.g., \textit{Put box into the drawer}, where the box is glued and thus cannot be picked up). Please refer to our project website\footref{projectwebsiteref} to view all the prepared setups, corresponding object properties, and instruction sentences.

Given these setups, we recorded the following datasets:

1) $\mathbf{D_{\mathcal{A}}^{real}}$: Aligned dataset with the same instructions given by both gestures and language, i.e., the user atempts to convey the valid instruction both through gestures and language. This dataset consisted of $20$ samples (one sample for each of the setups) for each user, i.e., $60$ samples in total.

2) $\mathbf{D_{\mathcal{U}}^{real}}$: Nonaligned dataset with a mismatch between the target action or objects in language and gesture instructions. The correct action and object can be determined either based on the properties of the object or by number and type of action parameters. For each of the $20$ setups, we recorded $4$ samples (i.e., each of the two alternative instructions was given either by language or gestures, while the other modality gave the valid instruction). This dataset consisted of $80$ samples for each user, i.e., $240$ samples in total.

\section{Experimental results}

First, we perform an ablation study (Sec.~\ref{sec:experiment-ablation-study}) to show the importance of different parts of our proposed system.
Secondly, we study the effect of different noise levels (Sec.~\ref{sec:experiment-noise-levels}), different merging methods, and thresholding approaches (Sec.~\ref{sec:experiment-thresholding}) on the model performance. We show the results both for the simulated setup and for the real datasets, where $3$ participants tested the proposed system.

\subsection{Ablation study}
\label{sec:experiment-ablation-study}

\begin{figure}[ht!]
  \vspace{0.8em}
  \centering
  \includegraphics[trim={0.4cm 0.4cm 0 0.4cm},width=0.8\linewidth]{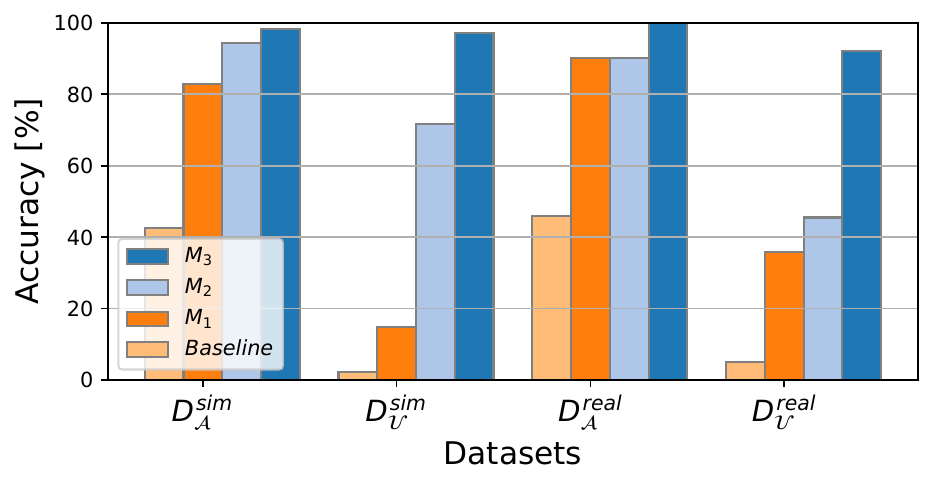}
  \vspace{-0.4em}
  \caption{Ablation study shows perfomance of the proposed model ($M_3$) compared to models without individual penalization functions ($M_2$, $M_1$) and towards the baseline. The baseline corresponds to the merging of modalities by $argmax$ function without any penalization terms. The results are shown for aligned ($D_\mathcal{A}^{sim}$) and unaligned ($D_\mathcal{U}^{sim}$) simulated datasets as well as on the real datasets ($D_\mathcal{A}^{real}$, $D_\mathcal{U}^{real}$). Models $M_1$, $M_2$, and $M_3$ used \textit{add} merging function and entropy thresholding. Real noise $n^{real}_1$ was added to both simulated datasets.}
  \label{fig:exp01}
  \vspace{-1.5em}
\end{figure}

The first experiment compares models $M_1$, $M_2$, and $M_3$ (see Sec.~\ref{sec:models} with the baseline method and shows the effect of individual components of the proposed model $M_3$. For this comparison of models we used \textit{mul} merging function, real added noise ($n^{real}$), and entropy thresholding. In Fig.~\ref{fig:exp01} we can see that for the aligned datasets ($\mathbf{D_{\mathcal{A}}^{sim}}$ and $\mathbf{D_{\mathcal{A}}^{real}}$) all the models $M_1$-$M_3$ perform very well ($M_3$ still outperforming the other ones), showing ability to deal with this level of noise thanks to merging function. On the contrary, the baseline method that uses \textit{argmax} to merge data from gesture and language already drops the performance only to $42.6\%$, resp. $45.8\%$. In the unaligned datasets, where more context is needed for deciding on the intended action, we can see a significant accuracy drop for all models apart from model $M_3$, underlining the importance of individual components of the model in the case of noisy and misaligned inputs from different modalities. Similar performance, as well as accuracy drop, can be observed both for simulated and real data. 



\vspace{-0.4em}
\subsection{Noise influence}
\label{sec:experiment-noise-levels}

\begin{figure}[ht!]
  \centering
  \includegraphics[trim={0 0.4cm 0 0.4cm},width=0.8\linewidth]{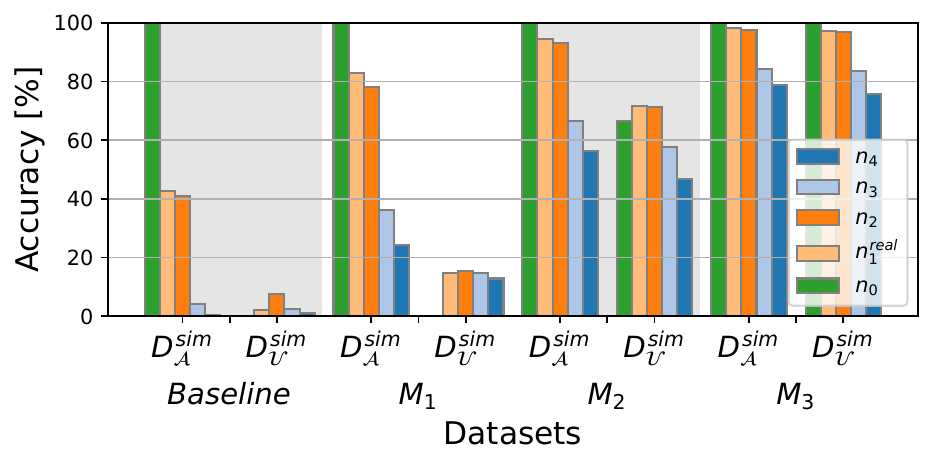}
  \caption{Accuracy scores on different noise levels ($n_{0, 1,2,3,4}$) for aligned and unaligned simulated datasets ($D^{sim}_{\mathcal{A}}$ $D^{sim}_{\mathcal{U}}$) for individual models. Results are shown for $add$ merging function and $entropy$ thresholding.}
  \label{fig:exp-noise}
\end{figure}

The second experiment shows noise influence (Sec.~\ref{sec:generating-noises}) on all models ($Baseline$, $M_1$, $M_2$, and $M_3$) with the best-performing merging settings (Merge function: $add$) and $entropy$ thresholding, see Fig.~\ref{fig:exp-noise}. 
Overall, we can see that the model $M_3$ is very robust to the noise, showing only a small decrease in accuracy with the increased noise. Furthermore, we can see for the model $M_3$ similar performance for the unaligned dataset. The real dataset included several samples with missing information in some of the modalities, however, the system correctly inferred in all of the cases the resulting action. 



\vspace{-0.2cm}
\subsection{Evaluation of merging approaches}
\label{sec:experiment-thresholding}

\begin{figure}[ht!]
\vspace{-1em}
  \begin{minipage}[c]{0.52\linewidth}
  \caption{The plot shows accuracy scores on different noise levels ($n_{0, 1,2,3,4}$) for aligned simulated datasets ($D^{sim}_{\mathcal{A}}$) for individual models. Results are shown for $add$ merging function and $entropy$ thresholding.}
  \label{fig:exp-noise2}
  \end{minipage}
  \begin{minipage}[c]{0.35\linewidth}
  \includegraphics[trim={0 0.0cm 0.0cm 0.0cm},height=1.0\linewidth]{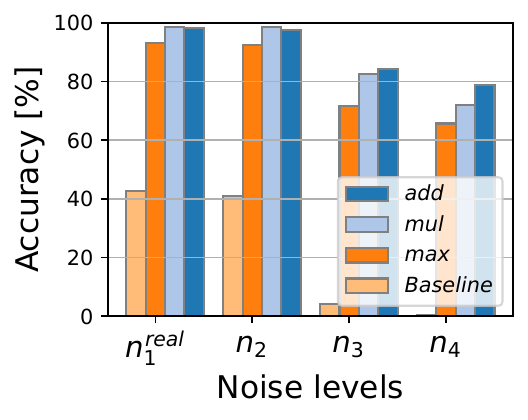}
  \end{minipage}
  \vspace{-2em}
\end{figure}

Finally, we evaluate different merging approaches, i.e. \textit{max}, \textit{mul}, and \textit{add} (see Tab.~\ref{tab:models}) for the full model $M_3$ with both penalization terms. In Fig.~\ref{fig:exp-noise2} we show the robustness of the approaches towards the noise. You can see that for the zero and low noise ($n_0$, $n_1^{real}$, and $n_2$), both \textit{mul} and \textit{add} perform similarly, outperforming significantly \textit{max} function. However, for higher noises \textit{add} outperforms the \textit{mul} merging approach. In other words, that \textit{add} merging function is more robust towards the noise.

\subsection{Evaluation of thresholding approaches}
\begin{figure}[ht!]
  \vspace{-1.2em}
  \centering
    \includegraphics[trim={0 0.5cm 0 0.5cm},width=0.9\linewidth]{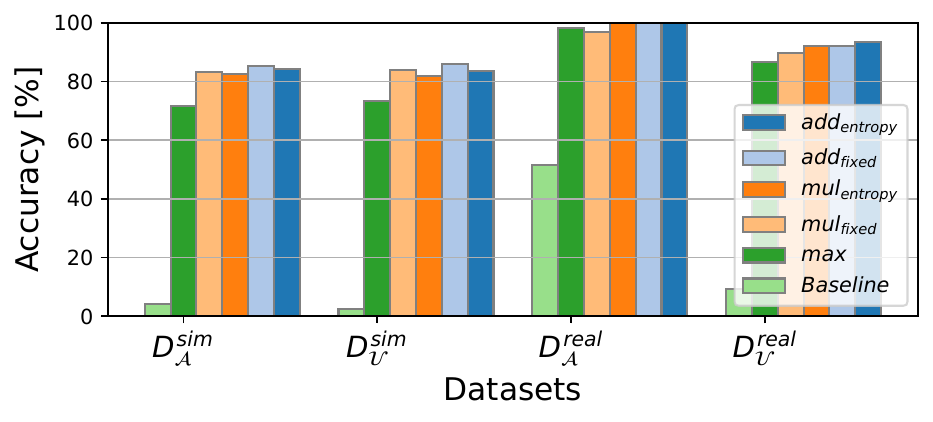}
  \caption{Thresholding approaches. Comparison of performance of fixed vs. entropy thresholding on both simulated (with noise $n_3
  $) and real datasets for model $M_3$.}
  \label{fig:plot5}
  \vspace{-2.2em}
\end{figure}

Finally, we evaluate fixed vs. entropy thresholding. We compare fixed thresholding which had thresholds manually tuned in the simulated environment with adaptive entropy thresholding. As expected, in simulation the fixed thresholding outperforms entropy thresholding, however, the difference is rather small (see Fig.~\ref{fig:plot5}, $add_{fixed}$ vs. $add_{entropy}$ for $D_\mathcal{A}^{sim}$, $D_\mathcal{U}^{sim}$). However, when the same fixed thresholds tuned for simulation were used in the real setup, the entropy thresholding already outperformed the fixed thresholding. This underlines the benefits of adaptive entropy thresholding, which can work well under different conditions such as different users, or tasks without manual tuning of thresholds.

\section{Conclusion}
In this paper, we proposed an approach for context-based merging of multiple modalities for robust human-robot interaction. Firstly, we evaluated the proposed method on several artificially generated bimodal datasets (however the method itself is not restricted to two modalities) with a controlled amount of noise,  as well as misalignment between data from different modalities. We show that the proposed model that takes into account the parameters of the actions and object properties is significantly more robust towards the added noise (reaching almost 76\% accuracy for highest noise, while the model $M_1$ which does not consider the context of the situation reaches only 24.3\% accuracy) (see Fig~\ref{fig:exp-noise}). 

Secondly, we evaluated the method on the real setup and performed a user study with 3 participants, each performing $100$ tasks in different experimental setups. The experiment contained tasks where commands from both modalities were aligned as well as tasks where one modality was providing misleading information and the system had to resolve the ambiguity by taking into account context of the situation. This experiment enabled us to also validate our simulated experiments. As can be seen in Fig.~\ref{fig:exp01} the results on the real dataset are similar to the simulated ones both for aligned and unaligned datasets.

Finally, we introduce an adaptive entropy-based thresholding method that enables automatic detection of thresholds between various interaction modes (i.e., detecting when to execute the action and when to query the user), yielding similar accuracy as the fixed thresholds in the simulation and better performance for the real setup (see Fig.~\ref{fig:plot5}). This enables easy adaptation of the method to new environments and scenarios, without the need for manually tuning and setting the thresholds.

\vspace{-0.8em}



\bibliographystyle{IEEEtran}
\bibliography{root}


\end{document}